# Assessing Refugees' Integration via Spatio-temporal Similarities of Mobility and Calling Behaviors

Antonio L. Alfeo, Mario G. C. A. Cimino, Bruno Lepri, Alex 'Sandy' Pentland and Gigliola Vaglini

*Abstract* — In Turkey the increasing tension, due to the presence of 3.4 million Syrian refugees, demands the formulation of effective integration policies. Moreover, their design requires tools aimed at understanding the integration of refugees despite the complexity of this phenomenon.

In this work, we propose a set of metrics aimed at providing insights and assessing the integration of Syrians refugees, by analyzing a real-world Call Details Records (CDRs) dataset including calls from refugees and locals in Turkey throughout 2017. Specifically, we exploit the similarity between refugees' and locals' spatial and temporal behaviors, in terms of communication and mobility in order to assess integration dynamics.

Together with the already known methods for data analysis, we use a novel computational approach to analyze spatio-temporal patterns: Computational Stigmergy, a bio-inspired scalar and temporal aggregation of samples. Computational Stigmergy associates each sample to a virtual pheromone deposit (mark). Marks in spatiotemporal proximity are aggregated into functional structures called trails, which summarize the spatiotemporal patterns in data and allows computing the similarity between different patterns.

According to our results, collective mobility and behavioral similarity with locals have great potential as measures of integration, since they are: (i) correlated with the amount of interaction with locals; (ii) an effective proxy for refugee's economic capacity, thus refugee's potential employment; and (iii) able to capture events that may disrupt the integration phenomena, such as social tensions.

## I. INTRODUCTION

IN the context of Syrian civil war, Turkey provides protection to over three million refugees [1]. Most of them were accepted in the early phase of the Syrian civil war when Turkey adopted an "open door" policy, i.e. do not ask for burden sharing, and avoiding the securitization of refugees. By means of this choice, Turkey was proving its role as a pivotal state in the Middle East, by actively contributing to the solution of humanitarian and political problems [2]. However, the "open door" policy was eventually revised, with a specific emphasis on internalization, burden-share, and camps within the border of Syria. In [2] such transition is associated with the concerns on the economic burden, the border security, the isolation in the international community, and the realization of false assumptions about the length of the crisis.

Indeed, in spite of the change of policy, the magnitude and the duration of the humanitarian crisis are resulting in an increasing tension in the local Turkish communities against the refugees. During the last years, many attempts to properly handle this situation have been done. The last one was the EU-Turkey deal, described by Amnesty International [3] as 'a disaster'. Thus, in order to prevent the growing of societal tensions over Syrian refugees, it urges the formulation of an effective long-term integration policy [4] [5].

However, the formulation of an effective policy demands tools for evaluating and understanding the integration of refugees. In this context, great benefits can be provided by complementing the paper-and-pencil surveys, the interviews, and the focus groups with big data-driven indicators [6] [7]. As an example, authors in [8] proposed a data-driven approach resulting in a significant improvement of refugees' integration.

Among the many possible sources of data on human behavior, great potential is offered by mobile phones [9], due to their wide penetration rate, i.e. 77% worldwide and 68% in developing countries [10]. Moreover, mobile phones can provide a variety of informative attributes, ranging from calling and texting activities to user's locations. Among these, Call Detail Records (CDRs) are a type of data collected by mobile telecommunication service providers, containing for each call/text the time, the location, the duration and the IDs of the users involved. By processing them it is possible to improve our understanding of complex and fast-changing situations and even drive an effective response in a variety of contexts, such as epidemic crisis [11] [12], crowd control problems [13], public violence [14], or refugee crisis [15].

In this work we analyze CDR datasets [16] to unfold the conditions that can contribute to the integration of refugees by studying their calling behavior and mobility. To this end, we employ a novel computational technique aimed at unfolding and matching spatio-temporal patterns, i.e. *Computational Stigmergy*. As a final result, we propose a set of metrics able to support policy makers while measuring the effectiveness of their integration policies.

In the following sections we present our approach and the results obtained by analyzing these data. In section 2 we present a literature review about the approaches based on mobility and behavioral analysis with a focus on the migratory

Antonio L. Alfeo , Mario G. C. A. Cimino and Gigliola Vaglini are with the Department of Information Engineering, University of Pisa, (largo Lucio Lazzarino 1, Pisa, Italy). Email: luca.alfeo@ing.unipi.it, mario.cimino@unipi.it, gigliola.vaglini@unipi.it. Bruno Lepri is with Bruno Kessler Foundation, via S. Croce, 77, Trento, Italy. Email: lepri@fbk.eu. Alex 'Sandy' Pentland is with the Massachussetts Institute of Technology. Email: pentland@mit.edu

phenomena. In section 3 we describe the proposed approach, whereas in section 4 the experimental setup and the analyzed data are presented. The obtained results are discussed in section 5. Finally, we draw the conclusions of this study in section 6.

## II. Related Works

Policy makers and humanitarian organizations have always relied on surveys to evaluate the effects of migrations and the effectiveness of migration policies. As an example, authors in [17] study the effect of the exposure to refugees on political outcomes in Hungary, by means of an approach based on surveys. As highlighted by the authors, the major limitation of such a method lies in the risk of biased answers since the surveys are self-reported and recorded after the crisis. Due to limitations like these, more and more studies are employing a big-data driven approaches [18]. As an example, in [19] authors exploit anonymized Facebook data to assess the assimilation of Arabic-speaking migrants in Germany by considering interests' similarities between migrants and host population. In [20] authors analyze the effect of the inflows of Syrian refugees in Turkey on political elections. In spite of the polarized attitude towards refugees, the presented results suggest that the inflow of refugees slightly reduced the support for the pro-refugees party.

In addition to a better understanding of migration effects, a data-driven approach can provide better insights into the effectiveness of migration policies. With such approach, authors in [21] were able to simulate the migration policies' implementation, detecting some counter-intuitive effects, e.g. some restrictive policies do not reduce the migratory flows, whereas they simply increase the illegal ones.

In this context, great potential has been shown by the analyses based on mobile phone data. As an example, these data are exploited in [22] to study how Syrian and Iraqi refugees used smartphones to reach and settle down in Turkey. In [23] authors use mobile phone data to assess the segregation of individuals according to their mobility in the urban area and their demographic characteristics. In contrast to assimilation theory, results show that the spatial behavior of the minority group has not become similar to that of the counterpart over generations.

Clearly, in the analysis of the migration phenomenon as a whole, it is not possible to ignore its changes over time, both in terms of the behavior of migrants and in terms of the impact of the migration. Indeed, the change in the phenomenon, in the policies, and in the response of migrants to the policies can be difficult to ascertain from past observations [24]. For such cases, approaches that model the underlying processes by taking into account the migrants' behavior over time are required.

This is confirmed by a number of studies, in which the inclusion of behavioral analysis results in a significant improvement of the quality of the predictions. This is the case of [25] in which by modeling the behavior of an individual, both in space and time, the authors obtain an improvement of the prediction performance up to 49% with respect to the demographic models. Again, in [26] the analysis of individuals' behavior over time is employed to detect events via mobile phone data. Authors conclude that the same analysis can be provided by addressing the mobility patterns in order to have a complete picture of the phenomenon.

More specifically, some recent works have started analyzing migrations and interactions at the community level by exploiting people's collective and individual mobility [27]. This is the case of [28] in which the authors employ geo-referenced data for about 62,000 individuals to estimate a set of US internal migration flows. Again, in [29], researchers analyze the behavior of migrants from the Middle East and Northern Africa to Europe by considering their collective mobility and its evolution in time. They highlight that one of the main problems in their study is the scarcity of geo-tagged tweets, resulting in difficulties in unfolding refugees' spatiotemporal patterns.

In this context, mobile phone data can offer more potential [30]. By exploiting them, many approaches aimed at analyzing human mobility can be derived [31]. At an individual level, authors in [32] achieved promising results in predicting individuals' trajectories by analyzing their mobility entropy. Other measures used with individual mobility are net capacity [33], radius of gyration [34], longest common subsequence (LCSS), and Fréchet distance [35] to name a few. The interested readers can refer to [31] for an extended survey of these approaches. However, in this work, we address collective mobility (i.e. made by groups) rather than individual one. Of course, the latter can also be obtained by aggregating the results obtained with the former, e.g. via a clustering procedure. Still, using an approach designed for the analysis of group mobility can be more convenient as it avoids two modeling phases, i.e. individual's mobility models and their aggregation to model the mobility of the group.

The techniques aimed at addressing collective mobility can be grouped into 2 main classes [36]: gravity models and intervening opportunities models. The gravity model considers population and distance as incentives and costs in determining mobility. The intervening opportunities models introduce as an incentive to mobility the number of opportunities (e.g. places of interest) between the origin and the destination. As an example, authors in [37], employ a kernel density estimation approach to approximate the underlying data distribution and a gravity model that generates flow maps; however, this approach seems to be inappropriate for complex, recurrent and dense flow patterns.

In general, these approaches (i) rely on origin-destination matrices, neglecting the dynamics occurring between the collective mobility patterns (e.g. temporal consistency, group's cohesiveness or sparsity) and (ii) are affected by a poor representation of the temporal dynamics, often exploited as an additional dimension of the spatial mobility patterns [38]. In this work, both issues are addressed with an approach based on Computational Stigmergy [39], which intrinsically embodies the time domain and unfold spatiotemporal dense mobility patterns [40].

## III. PROPOSED APPROACH

In order to assess the integration of refugees, it is essential to establish metrics able to capture this phenomenon. The metrics proposed in this work are based on the belief that integration is a process that allows the inclusion of an individual within a community through socialization and assimilation of practices and habits [41]. In other words, the integration of refugees could be measured according to the similarity of their behavior with the locals.

We propose the following metrics with the aim of exploiting a CDR dataset to gain insights about the integration of refugees through their behavioral and mobility similarity with locals.

*Refugee's Interaction Level* (IL), it is defined as the percentage of phone calls toward locals ($calls_{r \rightarrow L}$) made by a given refugee ($r$) in a given period of time. It represents how much the refugee is socially connected to the local community [10], i.e. 0 means no calls toward locals and 1 means only calls toward locals (see Eq. 1). Each level is defined as a range of 20% within this scale (i.e. 0-0.2, 0.2-0.4, 0.4-0.6, 0.6-0.8, 0.8-1). In our work, as in other previous studies in this field [42], we consider the IL a solid metric for measuring individuals' social integration.

$$IL_r = \frac{|calls_{r \rightarrow L}|}{|calls_{r \rightarrow L}| + |calls_{r \rightarrow R}|} \quad (1)$$

*Refugee's Calling Regularity* (CR). We represent the individual's daily calling behavior by considering the time series of the calls' frequency (i.e. the *calling pattern*) made by each individual. Specifically, we build the *calling pattern* as the number of phone calls made by a person in a given hour of the day during a given period of time. We normalize this amount with the average number of calls per hour in order to be comparable despite the different amount of calls made by each person.

In general the calling pattern may be due to several reasons, e.g. daily routines, habits, or working schedule. Even if it is not possible to determine which component has a predominant role in generating a specific calling pattern, we can assume that similar routines will most likely generate similar calling patterns. Moreover, routines' similarity is often linked to integration [43] [44]. Thus, the similarity between the calling patterns of locals and refugees may be considered as a proxy of integration.

For example, each refugee who is employed is supposed to have a daily routine (thus, a calling pattern) similar to the average calling pattern of the locals, since they are mostly employed [45]. In this context, the more a refugee's calling pattern $CP_r$ is similar to the average local's calling pattern *LCP* the more it is considered regular.

We employ the *cosine similarity* (Eq. 2) to compare 2 calling patterns due to its interpretability and robustness toward missing values [46], which commonly happen with sparse data such as CDR. Indeed, many refugees are characterized by a low number of phone calls. When those are also sporadic (i.e. happen only in a few hours of the day) the resulting calling pattern may be characterized by a number of missing values, which are ignored by the cosine similarity. This metric is defined between 0 (completely different calling pattern w.r.t. locals) and 1 (identical calling pattern w.r.t. locals).

$$CR_r = \frac{CR_r \cdot LCP}{||CR_r|| \cdot ||LCP||} \quad (2)$$

*Refugee's Mobility Similarity* (MS). Refugees' integration may result in sharing the same urban space at the same time with the host community [47] [48]. In order to measure such phenomenon by using call data we can collect the time and locations of each call occurred during the day to model the daily users' mobility. The similarity between refugees' and locals' mobility patterns can be computed by using the principle of stigmergy [49].

Stigmergy is a self-organization mechanism used in social insect colonies [50]. Basically, individuals in the colony affect each other behavior by marking a shared environment with pheromones when a specific condition occurs (e.g. the presence of food). The pheromone marks aggregate with each other in the trail if they are subsequently deposited in proximity to each other, otherwise they evaporate and eventually disappear. Thus, areas in which the condition above (e.g. the presence of food) occurs consistently are marked with a stable pheromone trail, which steers the colony toward the source of food. This pheromone-like aggregation mechanism is reproduced by Computational Stigmergy to unfold consistent spatio-temporal patterns in data [51] which are summarized by the virtual pheromone (i.e. stigmergic) *trail* [52].

By employing Computational Stigmergy, each call is transformed in a virtual pheromone mark and released in a virtual environment in correspondence to the call's location and instant in time (Fig. 2c). Moreover, at each time instant, the trail is subject to an evaporation process, i.e. a temporal decay with rate δ (Fig. 2b). The evaporation can be counteracted only by aggregating (Fig. 2d) new marks, i.e. summing them up to the existing stigmergic trail and reinforcing it. In a nutshell, the trail appears and stays only in correspondence of consistent marks depositing, i.e. spatio-temporal dense patterns in the data, summarized by the trail itself [53]. Eq. 3 describes the trail at time instant $i$.

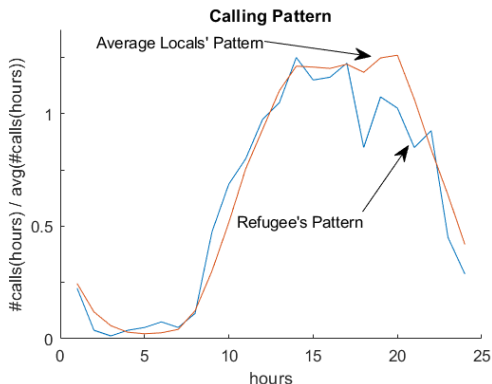

Fig. 1. Calling patterns in an average day. The figure represents the average number of calls for each hour, normalized with respect to the average number of call per hour. In orange we depict the average calling pattern of the locals, whereas in blue we present the calling pattern of one refugee.

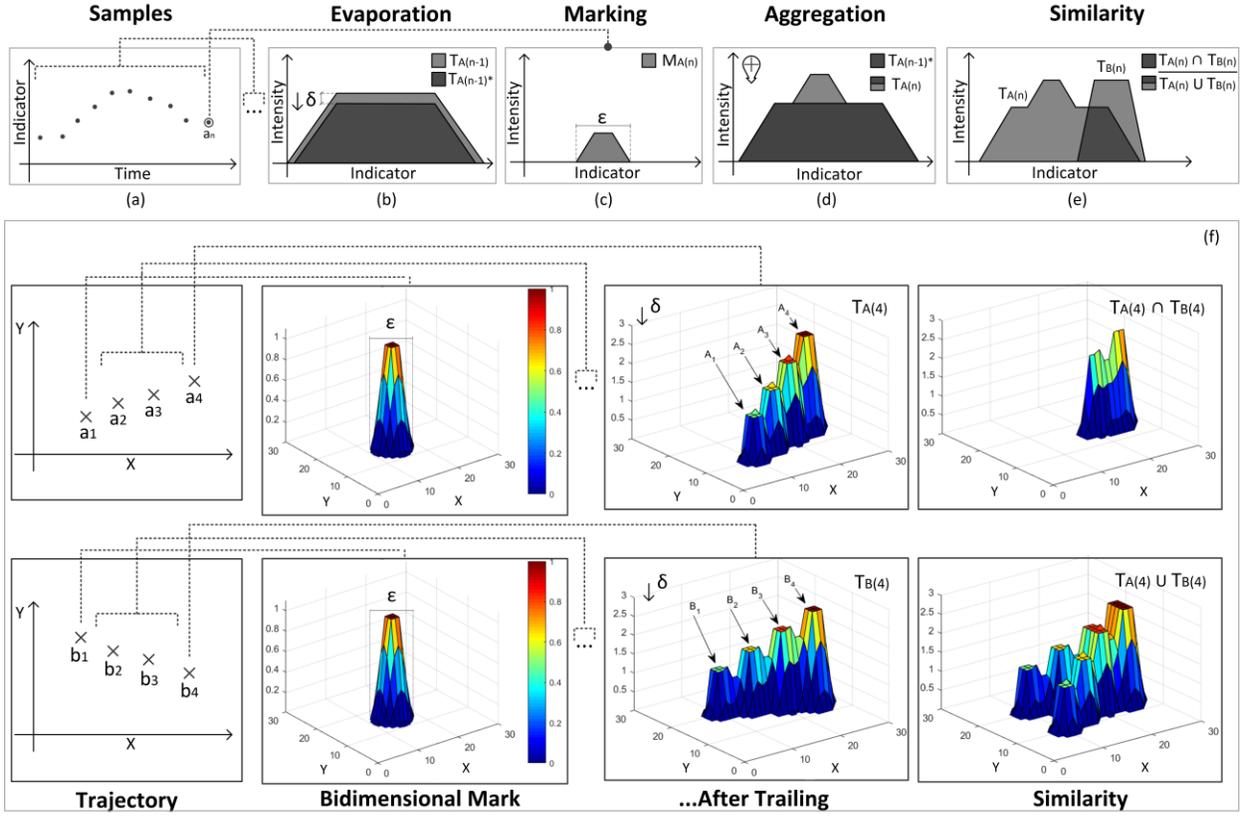

Fig. 2. Architecture based on Computational Stigmergy to measure the mobility similarity. Illustration of the samples processing modules (a-e), and its application to the comparison of 2 simple trajectories (f).

$$T_i = T_{i-1} - \delta + \text{Marks}_i \qquad (3)$$

It follows that the virtual pheromone mark and the evaporation are the main components of the stigmergic mechanism. Indeed, (i) the mark's width defines the proximity within two marks that can sum up with each other, meaning that two samples within this distance are considered in the same urban area, and (ii) the evaporation provides a memory mechanism to the stigmergic space as it results in the maximum time a mark can remain in the virtual space and contribute in building the trail. In our application context, each sample is transformed in a tridimensional mark that has the shape of a truncated cone (Fig. 2f), a width corresponding to 1 km in the real world scenario, and an evaporation rate $\delta$ set to 10%.

In this work, we use the trail to model collective mobility patterns, thus we build the trail by employing the samples of a group of users, e.g. refugees with a given IL. By matching different trails (Fig. 2e), we provide a general similarity measure for those patterns. The similarity between stigmergic trails is obtained by using an extended version of the Jaccard similarity [54], aimed at comparing trails as the ratio of their intersection (minimum common volume) and union (surfaces' maximum volume), as shown in Fig. 2f. This similarity is defined between 0 (completely different) and 1 (identical).

Therefore, the *mobility similarity* (MS) of a group of refugees is defined as the Jaccard similarity of the stigmergic trail obtained with the samples of this group and the stigmergic trail obtained with an equally sized group of locals. For the sake of understandability, in Fig. 2f we summarize the phases of the *mobility similarity* (MS) computation with 2 synthetic sequences of calls. We represent the trails ($T_{A(4)}$ and $T_{B(4)}$) obtained from the deposit of 4 consecutive samples ($A_1$, $A_2$, $A_3$, $A_4$ and $B_1$, $B_2$, $B_3$, $B_4$) of the synthetic sequence (A and B), their intersection and their union, which are used to compute their similarity. In our investigation, the trails to match are the ones obtained with groups of refugees $T_R$ and locals $T_L$ (Eq. 4). As an example, in Fig. 3 we show the trails obtained with a group of locals and two different groups of refugees. For those, we report the resulting MS.

$$\text{MS}_{R,L} = \frac{|T_R \cap T_L|}{|T_R \cup T_L|} \qquad (4)$$

IV. EXPERIMENTAL SETUP

In this section, we detail the data used in this study, its selection and preprocessing, and the further metrics used in our analysis.

A. Data description

In this work we analyze three different CDR datasets [16] provided by Turk Telekom in the context of the research challenge Data4Refugee (D4R).

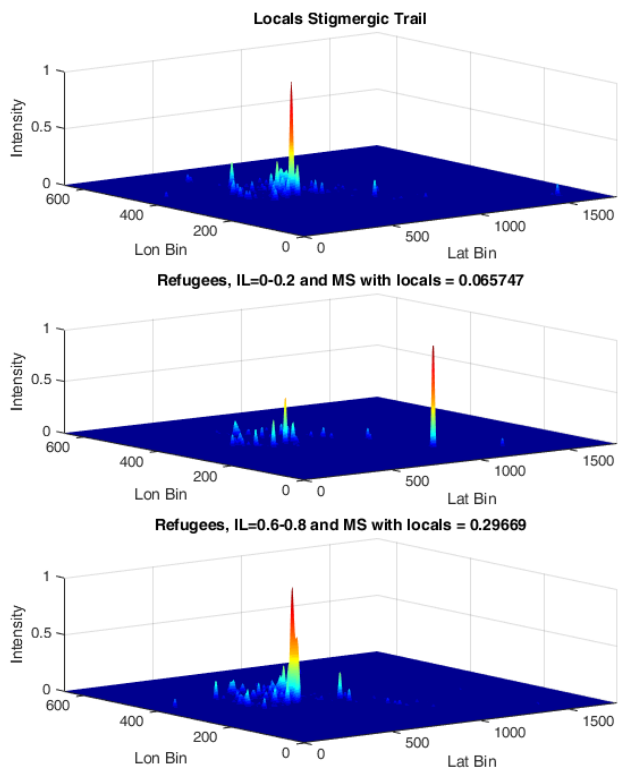

Fig. 3. Example of mobility similarity computation with two groups of refugees with different IL. The samples (calls) of locals are subsequently aggregated and evaporated forming the trail that summarizes their collective mobility pattern in December 29, 2017. This procedure is repeated with each group of refugees. At the end the obtained trails are compared with the locals' one.

In order to provide the reader with some insights about each D4R dataset, we briefly present each one of them:

a) The *Fine Grain Mobility Dataset* (FGMD) contains the antenna identifiers used by a group of randomly chosen users for each period of 2 weeks, for a total amount of 26 periods during 2017. This data has been anonymized by replacing the user number with a random ID, which prefix means refugee (i.e. 1), non-refugee (i.e. 2), and unknown (i.e. 3). In this dataset we can find, for each call, the caller id, the timestamp, the callee prefix, and the antenna id. Specifically, the call records collected in the dataset are linked to 600,000 refugees' id, 2.770.000 locals' id, and no "unknown" user's id. On the other hand, the number of calls toward refugees (locals) is about 2,082,000 (149,238,000). Finally, the number of calls with a callee prefix of type "unknown" (i.e. 3) covers just 0.93% of the whole dataset and they are dropped from our analysis, as they can be considered as noisy data [16].
b) The *Coarse Grain Mobility Dataset* (GGMD) contains the calls details for a unique group of users throughout the whole year, but with a more coarse spatial granularity, i.e. the district. In this case there is not reference to the callee, while the caller can be identified as refugee or local according to the prefix of their caller id (same as FGMD).
c) The *Antenna Traffic Dataset* (ATD) contains one year site-to-site traffic on an hourly basis. Each site is an antenna with known GPS location. Specifically, for each antenna there is a timestamp, the outgoing antenna, the incoming antenna, the total number of calls, the total number of refugees' calls, the total calls duration, and the total refugees' calls duration.

These datasets can be exploited with different aims and according to different analysis features, such as the number of individuals (single or groups), the spatial granularity (antenna-wise or district-wise) and the time window (daily, weekly, biweekly or monthly).

As an example, the information in the FGMD can be used to define the mobility patterns of refugees and locals, or to assess the interactions (calls) of refugees and locals. Thus, this information can be used to compute the *interaction level* and the *mobility similarity* on a bi-weekly base. On the other hand, the information contained in the CGMD can be used for long-term analysis (several months or year-round), and district-wise patterns, for analyzing the difference in call patterns. For this reason, the CGMD is employed for (i) the definition of additional district-wise metrics (section 4.B), and (ii) the computation of the *calling regularity*. It is worth noticing that both CGMD and FGMD are subjects to a selection process motivated and detailed in section 4.C.

Finally, the data contained in the ATD are used to gain some knowledge about the spatial distribution of refugees and their calling activity, since this data takes into account the whole population of refugees at a community level. Based on this knowledge an additional metric is proposed in section 3.B.

*B. Additional metrics*

In order to validate our approach and have an even more complete view of the phenomenon of refugees' integration, it is necessary to understand the circumstances that can foster or prevent their integration. For this purpose, we propose a further set of metrics focused on the analysis of those local characteristics (district-wise) that can influence the integration process, such as the cost of living, or the presence of other refugees in the same place.

*Residential Inclusion by District* (RI): We assume that most of the calls during the night and early morning come from people's houses. Indeed, based on this assumption many works in the field of the CDR analysis infer the location of an individual's home as the place from which they mostly calls between 8 pm and 8 am [55] [56]. Thus, by observing the percentage of calls made by refugees (via the ATD dataset) between 8 pm and 8 am per antenna $a \in d$ is possible to assess the coexistence of resident locals and refugees in a given district $d$ and a given month $m$. This metric is defined (see Eq. 5) between 0 (no resident refugees' in the district) and 1 (only resident refugees' in the district).

$$RI_{d,m} = \frac{|\text{calls}_{a,m}(R)|_{a \in d}}{|\text{calls}_{a,m}(R) + \text{calls}_{a,m}(L)|_{a \in d}} \quad (5)$$

*District Attractiveness* (DA): As for the assumptions used with the RI metric, a refugee resides in a given district and month if that district is the most recurrent location from which they make calls between 8 pm and 8 am. A district is considered attractive according to the percentage of resident refugees who are not leaving the district in that month, i.e. reside in the same district also in the next month. Specifically, given *residentRef*$_{d,m}$ i.e. the set of the refugees who live in the district $d$ during the month $m$, the *district attractiveness* is defined by Eq. 6.

$$DA_{d,m} = \frac{|\text{residentRef}_{d,m} \cap \text{residentRef}_{d,m+1}|}{|\text{residentRef}_{d,m}|} \quad (6)$$

*District Cost of Living*: One of the main drivers of the social and economic integration of refugees is their employment condition. Unfortunately, it is not possible to have a clear picture of this, since they are often employed in the informal sector [57] [58]. However, it is possible to have some insights about this phenomenon by exploiting the characteristics of the location where refugees live. Indeed, depending on their economic well-being and the level of integration, refugees may choose different settlement solutions [59]. For example, only an individual who has enough economical resources (e.g. who has a job) can afford the *cost of living* in an area that offers better living conditions. On the other hand, those who are not integrated and/or not employed are more likely to be socially isolated and relegated to poor neighborhoods. For this reason our analysis addressing the socio-economical integration of refugees exploits the average rent cost per square meter in a given district in the year 2016 as an indicator of the *cost of living* for that specific district. The source of these data is a housing site that wants to remain anonymous.

*C. Data Selection*

Since our investigation includes an analysis of mobility and calling behavior, it is fundamental to focus on areas with (i) high calling activity made by refugees, i.e. avoiding regions with sparse or sporadic calls, which may result in unrepresentative behavioral models; and (ii) a good spatial resolution, which means high density of antennas, since the granularity of the mobility patterns is antenna-wise; indeed, in an area with few antennas all mobility patterns will be roughly similar; and (iii) an area with a great variety of living conditions (e.g. cost of living, services' availability) in order to consider the settlement choice of each refugee according to their social integration and economic condition. This information (e.g. cost of living, services availability) are usually collected by district. Thus, the observation of the settlement choice over time requires focusing on an area with many different districts close to each other. This allows refugees to move from one district to another according to the change of their conditions over time. For instance, it is unlikely that a refugee who is prone to segregation resides in a district where no other refugees reside as well as it is unlikely that an unemployed refugee can live in a neighborhood with a high cost of living, as they may have problems in affording it. However, if such conditions happen to change over time the district where a given refugee lives may change accordingly.

Therefore, we first aim at finding the areas with (i) high refugees' calling activity, (ii) high density of antennas, and (iii) high number of districts covering different living conditions (e.g. cost of living). We analyze the total amount of calls (in seconds) made by refugees (Fig. 4), and the density of antennas (Fig. 5) with a spatial discretization of 10 km per squared areas over the whole Turkey by exploiting the ATD dataset.

The cities of Istanbul, Ankara and Izmir are the most promising areas to conduct our analysis since they have the larger density of antennas and the larger calling activity made by refugees. This conclusion is also confirmed by the descriptive statistics of the D4R dataset provided in [16]. Moreover, Istanbul's metropolitan area consists of 69 districts [16] with a variety of different characteristics (e.g., different housing costs or job opportunities).

Moreover, we adopt a few countermeasures to address a common problem with CDR data analysis, i.e. data sparsity [60] [61]. Firstly, we select only the individuals in the dataset with at least 2 samples (i.e. calls) on average per day. Secondly, we group individuals' samples according to their *integration level* or time period. Specifically, in the mobility analysis, we group the samples of all users belonging to the same IL group, in this way we can use the samples of a number of users (thousands in Istanbul, see Fig. 9) to build collective mobility patterns. On the other hand, the individual's calling pattern is built by using all the samples in a given time period, i.e. 14 days (with FGMD) and one month (with CGMD), since we model the distribution of the calls during the day rather than its change from day to day in the same time period. As a result, the number of samples used for a calling pattern is an order of magnitude higher than the average number of samples in a day, but this is insufficient compared to the improvements guaranteed with the mobility analysis. For this reason, we focused the analysis addressing the *calling regularity* on Istanbul, where the average amount of refugees with more than 2 calls per day is 10 times greater than in Ankara and Izmir. Moreover, we have data on the districts' cost of living in Istanbul and therefore we can investigate the correlations between those and our metrics.

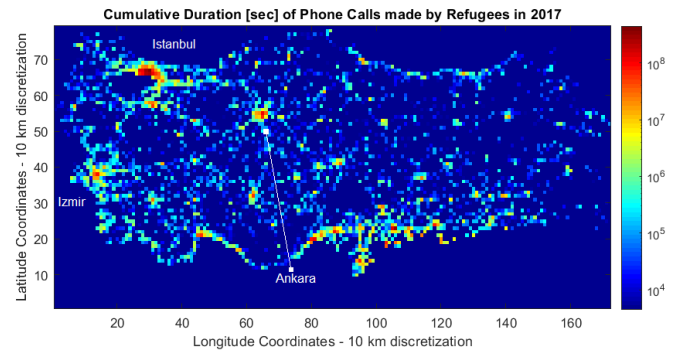

Fig. 4. Cumulative duration of calls during 2017 for each squared area in Turkey. Squared areas are 10km x 10km and drawn from a grid that we overlay on the map. We have highlighted the location of three major cities (Istanbul, Izmir and Ankara).

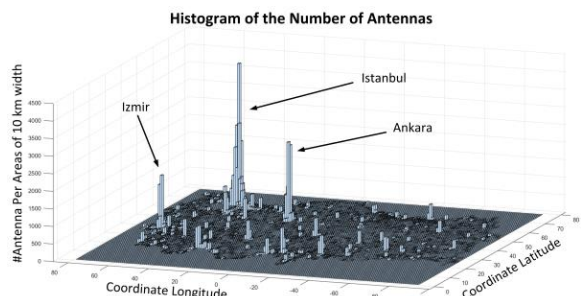

Fig. 5. Number of antennas per squared area in Turkey. The bars' height represents the number of antennas counted in each location. Locations are 10km x 10km and drawn from a grid that we overlay on the map. We have highlighted the location of three major cities.

Finally, in the analysis with DA, we consider the refugees who reside in Istanbul for at least half of the whole year, in order to mitigate the impact of the newcomers on the analysis. By excluding the districts in which none of the selected refugees is located, the total amount of districts analyzed in Istanbul is equal to 38.

## V. RESULTS

In this section, we describe and discuss the findings of our analyses obtained by employing the metrics and the data presented in the previous sections.

### A. Calling regularity and interaction with locals

In order to verify if the *calling regularity* can be used as an integration proxy, we analyze the relationship between it and the *interaction level* of each refugee in Istanbul. Specifically, we compute the Pearson correlation coefficient between the average *calling regularity* of the refugees in Istanbul and their *interaction level* in each of the 25 time periods. The distribution of the coefficients has the first, second and third quartiles equal to 0.76, 0.79, and 0.93, respectively. The resulting p-value has a 95% confidence interval equal to [0.06,0.25]. The *interaction level* and the *calling regularity* are positively correlated, providing us with the insight that refugees that exhibit greater interaction with locals may have also similar daily routines. This result confirms the findings in [43] and [44] and should be considered while designing the policies addressing refugees' integration.

### B. Districts, inclusion and calling regularity

We analyze the relationship between the *district attractiveness*, the *residential inclusion* and the *calling regularity* of the refugees in each district of Istanbul according to the *cost of living* in the district itself.

Firstly, we focus on the relationship between the *residential inclusion* and the *calling regularity*. Thus, we compute the correlation coefficients between the average *calling regularity* of the refugees living in a given district during a given month and the *residential inclusion* of the same district in the same month. Their distribution has the first, second and third quartiles fall to 0.45, 0.55, and 0.66, respectively. We take into account each district analyzed in Istanbul (i.e. 38) during each month of 2017 (i.e. 12). The resulting p-value has a 95% confidence interval equal to [0.07,0.18].

According to this result, the districts with higher residential inclusion may also be characterized by higher similarity between locals' and refugees' routines. This suggests that a minimum number of refugees per area is required for triggering the integration phenomenon, as suggested in [62].

In Fig. 6 we show the distribution of the *residential inclusion* by month and district in order to understand better these magnitudes. It is evident that many districts have a low RI, thus depicting a scenario of minor coexistence of refugees and locals. Moreover, in the few districts (and months) with higher RI, its value never exceeds 50%. Thus, the more evenly distributed the residents (locals and refugees in an area) are, the greater the similarity between the routines of locals and refugees.

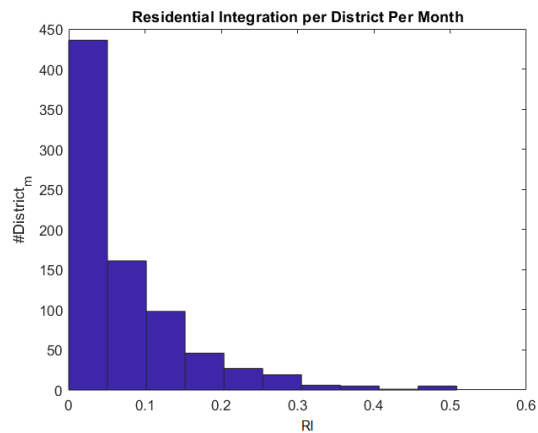

Fig. 6. Histogram of the distribution of RI per each month and district in Istanbul.

Moreover, we study if there exists a correlation between the average DA and the *cost of living* of each district in Istanbul (Fig. 7). With a coefficient equal to -0.56 and a p-value of 0.003, we conclude that DA and the *cost of living* are significantly and inversely correlated, which means that the cheaper is the *cost of living* the more attractive is a district for a refugee.

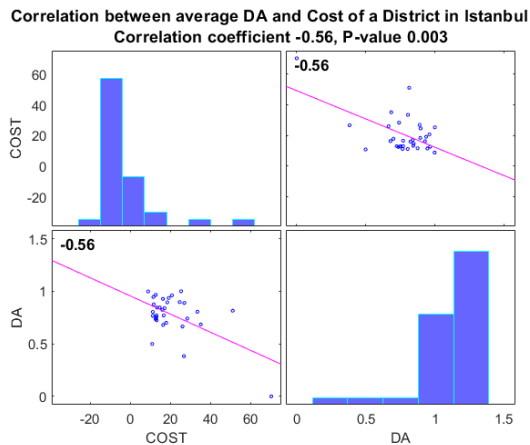

Fig. 7. Correlation matrix obtained with the yearly DA and cost of living for each one of the 38 districts analyzed in Istanbul. The correlation coefficient is -0.56 and the p-value is 0.003. On the diagonal the distribution of the average RI and the cost of living, the others are the bivariate scatter plots.

The results in the section 5.A show that the *calling regularity* can be considered a proxy for social integration. Moreover, as specified in section 3, the *calling regularity* can be linked to the employment of a refugee [63]. We can provide further insights in this context, by considering the characteristics of the location where refugees live [59]. Thus, we compute the correlation between the *cost of living* in each district in Istanbul and the average *calling regularity* of refugees living in that district (Fig. 8).

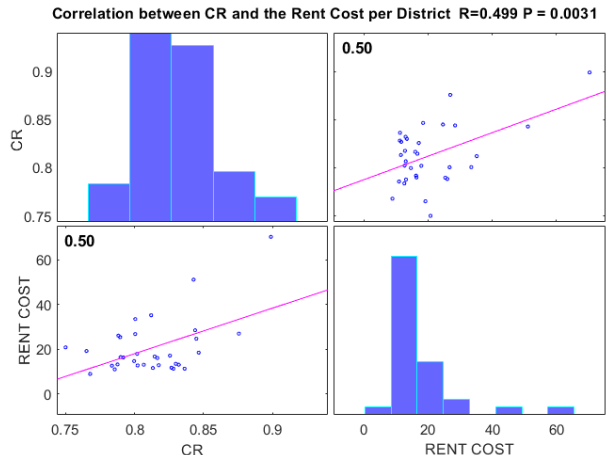

Fig. 8. Correlation matrix obtained with the average CR in a district and the cost of living for each one of the 38 districts analyzed in Istanbul. On the diagonal the distribution of the average CR and cost of living, the others are the bivariate scatter plots.

With a correlation coefficient equals to 0.5 and a p-value equals to 0.003, the average *calling regularity* of refugees living in the district exhibits a significant and positive correlation with the *cost of living* in that district. This means that, although it may be influenced by some factors not detectable by the data under analysis, the *calling regularity* is a proxy for the daily routine similarity and for the economic capacity of refugees (i.e. the ability to meet a certain *cost of living*), thus it is a tool able to capture the necessary conditions occurring together with the employment of refugees. This metric offers great potential in the analysis of refugees' integration, since having a job is one of the main steps in the integration process [64], but it is also hard to analyze it since the refugees' employment often happen in the informal sector [58].

This result is also confirmed by evaluating the correlation of the *cost of living* with the *distances* between calling patterns of locals and refugees, i.e. the opposite concept with respect to the *calling regularity*, which is based on their similarity. Indeed, according to our results (Table I), as the distance increases, the district's *cost of living* decreases. Moreover, our approach has proven to offer similar performances when compared against Dynamic Time Warping (DTW) [40] or Euclidean distance, two metrics widely used in time series analysis.

Finally, since we are dealing with spatial variables, we account for biases due to potential spatial autocorrelation. In order to evaluate such occurrence, we employ a spatial regression model aimed at predicting the average *calling regularity* of an Istanbul's district by using as predictors the normalized district-wise metrics presented in section 4.B, and as weight matrix the districts' normalized distances. Such normalizations are made via a min-max procedure.

TABLE I
CORRELATION DISTANCES BETWEEN LOCALS' AND REFUGEES' CALLING PATTERNS AND THE COST OF LIVING OF THEIR DISTRICT

| Distance Measure | Correlation Coefficient | p-value |
|---|---|---|
| Euclidean | -0.53 | 0.0015 |
| Cosine | -0.50 | 0.0031 |
| DTW | -0.53 | 0.0015 |

Moreover, we analyze the beta coefficients of the spatial regression in order to gain insights into the relationship between the metrics and the average *calling regularity*.

In this regard, the variables with the strongest correlation with the calling regularity are the *district attractiveness* and the *cost of living*. Both of them are characterized by a valuable statistical significance, which suggests how developing an economic capacity (e.g. being employed) and establishing a long-term residence provide the greatest contribution to the dynamics of integration. Moreover, by using this model we are able to provide a good approximation of the calling regularity, obtaining a Mean Square Error equal to 0.008. Finally, in order to assess potential misspecifications or biases in our model [65], we test if the achieved results and the residuals are spatially autocorrelated. According to the resulting Moran's I coefficients, we can assume that our study is robust toward the issues related to spatial autocorrelation. The above-discussed results are summarized in Table II.

TABLE II
SPATIAL REGRESSION MODEL BETWEEN CALLING REGULARITY AND DISTRICT CHARACTERISTICS (I.E. COST OF LIVING, DISTRICT ATTRACTIVENESS, AND RESIDENTIAL INCLUSION). THE BETA COEFFICIENTS ARE REPORTED IN THE TABLE. *P<0.01, **P<0.001.

| Name | Value |
|---|---|
| β Cost of living | 0.399** |
| β District attractiveness | 0.512** |
| β Residential inclusion | 0.148 |
| Mean Square Error | 0.008 |
| Moran's I predicted CR(p-value) | -0.056 (0.011) |
| Moran's I regression residuals (p-value) | -0.047 (0.071) |

### C. Mobility and Interaction with locals

Another fundamental driver of integration can be the sharing of urban spaces with the locals [66]. However, its positive contribution in the integration dynamics is not obvious. Indeed, it can allow the progressive integration of refugees in the social structure of the hosting city; or, on the other hand, the shared urban areas may be not perceived as a safe space [67] thus leading to the occurrence of social tension in those areas.

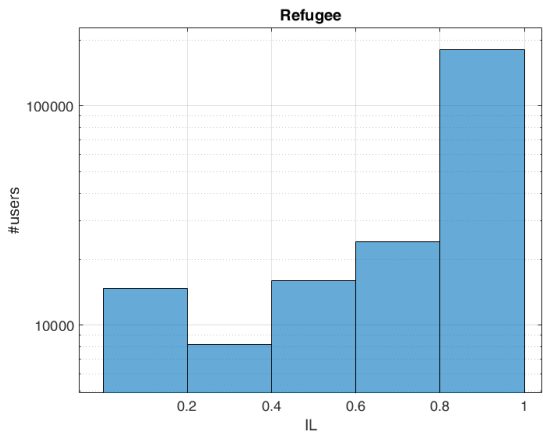

Fig. 9. Number of refugees in Istanbul according to the interaction level. Fine Grain Mobility Dataset. Log scale.

In order to understand the contribution of sharing the same urban space with locals, we analyze the relationship between *mobility similarity* and *interaction level* on a daily basis. Specifically, we create the collective mobility patterns of the group of refugees with a given *interaction level*, i.e. the stigmergic trails obtained with all the samples of the people in that group. Then, we compute the *mobility similarity* with the collective mobility patterns obtained with an equally sized group of locals. Regarding the size of these groups, we highlight that the *mobility similarity* measure is sensitive to the number of individuals employed in the creation of the collective mobility patterns, i.e. the more the individuals the higher the likely to have more similar collective mobility patterns. In addition to this, the size of the group of individuals with a given *interaction level* varies significantly (Fig. 9).

Thus, in order to have a fair comparison between the similarities computed with different groups, we set the size of each group as the minimum size among all the groups. Finally, we collect the Pearson correlation coefficients between the *interaction level* of each group and the resulting *mobility similarity*. We repeat this procedure multiple times by randomly subsampling the people for each group larger than the smallest one. In Fig. 10 we present the distribution of the obtained correlation coefficients by means of quartiles. In the 3 cities analyzed the *mobility similarity* is strongly correlated with the interaction level. Indeed, the distributions of the resulting correlation coefficients have a median equal to 0.91 in Istanbul, 0.82 in Ankara, and 0.92 in Izmir.

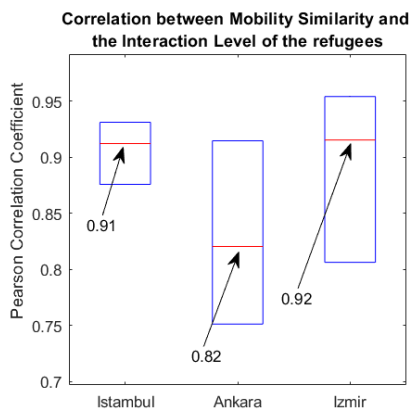

Fig. 10. Quartiles of the correlation coefficients between the mobility similarity and the interaction level of random groups (i.e. 5) of refugees with a given IL, over multiple (i.e. 5) trials. The p-value of the correlation coefficients have a 95% confidence interval equal to [0.004,0.01] in Istanbul, [0,0.091] in Ankara, and [0,0.64] in Izmir.

The more the refugees have interactions with locals, the more they share urban spaces with the locals. This allows us to claim that sharing urban spaces is a positive factor in the dynamics of integration of refugees. Thus, the policies designed to improve refugees' integration should take into account *mobility similarity* to assess their impact.

### D. Integration and Social Tension

By considering the correlation with the *interaction level* shown in section 5.C, the *mobility similarity* can be considered a proxy of refugees' integration. Thus, we exploit it to study the effects of the events that are certainly caused or can cause the disruption of refugees' integration: the occurrence of social tensions. In order to look for the features that characterize a social tension, it is necessary to start with few examples of publicly known social tensions.

Specifically, we collect a set of such events and we compare the *mobility similarity* and *interaction level* in 2 weeks before and after each event. We have found a number of occurrence of such events by searching for them over the internet [68] [69] and exploiting a publicly available news collector, i.e. the GDELT Project [70]. The GDELT Project monitors the world's broadcast, print, and web news from all over the world and makes it possible to query them according to locations, subjects involved, and emotions. By querying for events involving refugees in Turkey, we were able to obtain a pool of potential events that we checked manually to select only the ones related to actual social tensions and police interventions. The final pool of events taken into consideration is displayed in Table III.

TABLE III
DATES AND LOCATIONS OF THE SOCIAL TENSION EVENTS

| Day | Description | Location | Source |
| --- | --- | --- | --- |
| March 6 | Syrians caught by the police | Izmir | [86] |
| April 12 | Terrorist Attack | Istanbul | [87] |
| May 15 | Syrian Killed in a Brawl with locals | Istanbul | [88] |
| May 16 | Clashes between Syrians and locals | Istanbul | [89] |

Once these events have been identified, we study the impact of these social tensions by calculating the *mobility similarity* (with repeated trials according to the methodology described in the last section) and the percentage of calls made toward the locals, according to the *interaction level* of the refugees. These measures will be derived with data from different periods, by employing the FGMD. In order to make them comparable and highlight the fluctuations with respect to their average across different periods, a normalization with (i.e. divided by) their average amount per period is performed. Finally, we present the ratio between the *mobility similarity* and the percentage of calls in the two weeks before and after each event. If this ratio is greater than 1, it indicates that after the event the integration measure taken into consideration has decreased. As an example, in Fig. 11 we show the results obtained with the event of May 16 in Istanbul.

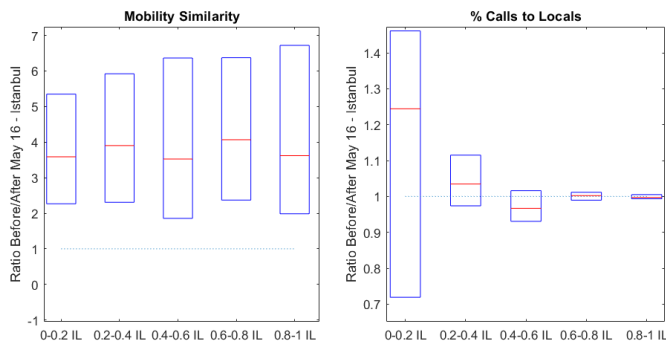

Fig. 11. Mobility similarity (left) and the percentage of calls made toward refugees (right): ratio between the values two weeks before and two weeks after the 16th of May in Istanbul. A ratio greater than 1 indicates that, after the event, the integration measure taken into consideration has decreased. The ratio is presented according to the IL of the group of refugees.

It is apparent that the social tension affects the behavior of the refugees by reducing the amount of shared urban space with the locals (i.e. lowering the *mobility similarity* after the event). Moreover, in terms of calls made toward locals, the social tension event has a greater effect on the group of refugees with lower *interaction level*.

Indeed, on average, they exhibit a lower percentage of calls made toward locals and a greater variability. Moreover, this trend is confirmed on every event we are taking into account, as shown in the aggregate results in Fig. 12. Indeed, the quartiles of the percentage of calls made toward locals are arranged as [0.55, 1.05, 1.41] with the refugees with the lower *interaction level*, whereas are [0.98, 0.99, 1] with the refugees with the greater *interaction level*. Moreover, the group of refugees with the lower *interaction level* exhibits the greatest decrease in *mobility similarity* after the social tension events. Indeed, the median of the distribution of the ratios obtained with the *mobility similarity* with the lower and greater *interaction level* are respectively 5.31 and 3, which means that the *mobility similarity* of the refugees with lower *interaction level* decreased 77% more with respect to the refugees with greater *interaction level*. Based on the obtained results, the proposed metrics seem able to capture the effect of a social tension and should be taken into account when addressing

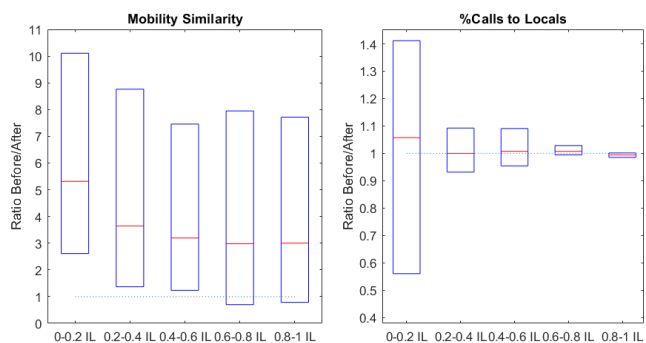

Fig. 12. Mobility similarity (left) and the percentage of calls made toward refugees (right): ratio between the values two weeks before and two weeks after each social tension. If this ratio is greater than 1, it indicates that the integration measure taken into consideration has decreased after the event. The ratio is presented according to the IL of the group of refugees.

applications such as attempting to identify or measure the impact of social tension events.

## VI. DISCUSSION AND CONCLUSION

In this work, we have proposed a set of metrics to assess the integration of Syrian refugees in Turkey, by considering the similarities with locals in terms of calling behavior and collective mobility.

Specifically, we found that (i) both *mobility similarity* and *calling regularity* are positively correlated with the *interaction level* between refugees and locals, and have proved to offer great potential as measures of the integration related phenomenon with different applications, (ii) the *calling regularity* is also a proxy for refugee's economic capacity, which may imply refugee's employment, (iii) the *mobility similarity* is affected by events such as social tension, and finally (iv) the behavior of less integrated refugees appears to be significantly more affected by this kind of events.

Yet, our findings may be limited by the representativeness of a behavioral model based on call data. Nowadays, there are many options for communicating other than calls, such as instant messaging platforms, e.g. WhatsApp [71]. Their usage may affect individuals' calling behavior, e.g. resulting in just occasional calls (i.e. sparse data samples) [72]. This problem is extensively addressed in [73], in which authors study individuals' social ties with CDR data. In this work, the authors acknowledge the goodness of call frequency as a metric for CDR data while advising caution in drawing conclusions, given that for example in their study several participants result to communicate almost exclusively using Facebook, whereas other participants extensively use email, Skype, or WhatsApp. Authors conclude that, even if a comprehensive behavioral model should consider those sources, including them may result in a titanic effort due to the following issue: (i) the increasing size and diversity of data to be analyzed as the number of communication channels continues to grow, (ii) each person may prefer different communication channels for different purposes, and switch between them according to fads or what it is used by their friends, (iii) linking individuals' identities across multiple communication sources is non-trivial, error-prone, and privacy-invasive. In the light of the above, the results of our work should be considered as very important insights about social integration within the growing CDR-based literature rather an exhaustive model to represent the integration phenomenon as a whole.

Still, our results can help drawing a few integration policy guidelines. Specifically, since the sharing of urban space with locals seems to be a valuable proxy for integration, policies that prevent the formation of ghettos and encourage the shared use of city areas should be considered in order to increase the integration. Moreover, an effective integration policy should address refugees' employment, since currently refugees seem more attracted to the less expensive districts, confirming a low employment issue, whereas the refugees who have the economic capacity to live in a more expensive district is likely to be more integrated. Finally, we have also shown that the group of more integrated refugees is more likely to be unrelated to social disruptive phenomena such social tensions, confirming the benefits of greater integration, which are well known in the literature.

Given the promising results obtained with these metrics, future work will focus on employing them (i) with datasets addressing other cities in order to further validate our results; and (ii) with a machine learning approach aimed at "nowcasting" the integration level of refugees [74], as it could result in a useful tool for policymakers.

ACKNOWLEDGMENT

Authors thank Turk Telecom and the collaborators (e.g. Bogazici University, Tubitak, Data-Pop Alliance, UNICEF, UNHCR, the International Organization for Migration, etc.) of the Data for Refugees Turkey (D4R) challenge as well as the Organization Committee and the Project Evaluation Committee. Authors also thank Marco De Nadai and Alfredo J. Morales-Guzman for their advices in the experimental phase and in the preparation of this manuscript.

REFERENCES

[1] E. F. Keyman, "Turkey at the heart of the refugee and ISIL crises: Can the buffer state be a solution," *Rev. Int'l L. & Pol.,* vol. 12, p. 5, 2016.

[2] S. N. E. G. Aras and Z. Mencutek, "The international migration and foreign policy nexus: the case of Syrian refugee crisis and Turkey," *Migration letters,* vol. 12, p. 193, 2015.

[3] G. Kondylia, "The EU-Turkey deal: Europe's year of shame," *Amnesty International,* vol. 20, 2017.

[4] A. Ager and A. Strang, The Experience of Integration: A qualitative study of refugee integration in the local communities of Pollockshaws and Islington, Research Development and Statistics Directorate, Home Office, 2004.

[5] E. Carpi and H. Pinar Senouguz, "Refugee Hospitality in Lebanon and Turkey. On Making 'The Other'," *International Migration,* 2018.

[6] K. Hardy and A. Maurushat, "Opening up government data for Big Data analysis and public benefit," *Computer Law & Security Review,* vol. 33, no. 1, pp. 30-37, 2017.

[7] M. Avvenuti, M. G. Cimino, S. Cresci, A. Marchetti and M. Tesconi, "A framework for detecting unfolding emergencies using humans as sensors," *SpringerPlus,* vol. 5, no. 1, p. 43, 2016.

[8] K. Bansak, J. Ferwerda, J. Hainmueller, A. Dillon, D. Hangartner, D. Lawrence and J. Weinstein, "Improving refugee integration through data-driven algorithmic assignment," *Science,* vol. 359, pp. 325-329, 2018.

[9] V. D. Blondel, A. Decuyper and G. Krings, "A survey of results on mobile phone datasets analysis," *EPJ data science,* vol. 4, no. 1, p. 10, 2015.

[10] J. Blumenstock and L. Fratamico, "Social and spatial ethnic segregation: a framework for analyzing segregation with large-scale spatial network data," in *Proceedings of the 4th Annual Symposium on Computing for Development*, 2013.

[11] A. Wesolowski, T. Qureshi, M. F. Boni, P. R. Sundsoy, M. A. Johansson, S. B. Rasheed, K. Engo-Monsen and C. O. Buckee, "Impact of human mobility on the emergence of dengue epidemics in Pakistan," *Proceedings of the National Academy of Sciences,* vol. 112, no. 38, pp. 11887-11892, 2015.

[12] M. Tizzoni, P. Bajardi, A. Decuyper, G. K. K. King, C. M. Schneider, V. Blondel, Z. Smoreda, M. C. Gonzalez and V. Colizza, "On the use of human mobility proxies for modeling epidemics," *PLoS computational biology,* vol. 10, no. 7, p. e1003716, 2014.

[13] Y. Dong, F. Pinelli, Y. Gkoufas, Z. Nabi, F. Calabrese and N. V. Chawla, "Inferring unusual crowd events from mobile phone call detail records," in *Dong, Y., Pinelli, F., Gkoufas, Y., Nabi, Z., Calabrese, F., & Chawla, N. V. (2015, September). Inferring unusual crowd events from mobile phone call detail records. In Joint European conference on machine learning and knowledge discovery in databases*, 2015.

[14] D. Gundogdu, O. D. Incel, A. A. Salah and B. Lepri, "Countrywide arrhythmia: emergency event detection using mobile phone data," *EPJ Data Science,* vol. 5, no. 1, p. 25, 2016.

[15] J. Qadir, A. Ali, R. ur Rasool, A. Zwitter, A. Sathiaseelan and J. Crowcroft, "Crisis analytics: big data-driven crisis response," *Journal of International Humanitarian Action,* vol. 1, p. 12, 2016.

[16] A. A. Salah, A. Pentland, B. Lepri, E. Letouzé, P. Vinck, Y.-A. de Montjoye, X. Dong and Ö. Da\ugdelen, "Data for Refugees: The D4R Challenge on Mobility of Syrian Refugees in Turkey," *arXiv preprint arXiv:1807.00523,* 2018.

[17] T. Gessler, G. Toth and J. Wachs, "No country for asylum seekers? How short-term exposure to refugees influences attitudes and voting behavior in Hungary," *SocArXiv,* 2019.

[18] C. Hughes, E. Zagheni, G. J. Abel, A. Sorichetta, A. Wi'sniowski, I. Weber and A. J. Tatem, "Inferring Migrations: Traditional Methods and New Approaches based on Mobile Phone, Social Media, and other Big Data: Feasibility study on Inferring (labour) mobility and migration in the European Union from big data and social media data," *EU publications,* pp. 1-41, 2016.

[19] A. Dubois, E. Zagheni, K. Garimella and I. Weber, "Studying migrant assimilation through facebook interests," in *International Conference on Social Informatics*, 2018.

[20] O. a. K. N. Altindag, "Do refugees impact voting behavior in the host country? Evidence from Syrian Refugee inflows in Turkey," 2017.

[21] M. Simon, C. Schwartz, D. Hudson and S. D. Johnson, "A data-driven computational model on the effects of immigration policies," *Proceedings of the National Academy of Sciences,* vol. 115, pp. E7914--E7923, 2018.

[22] K. Smets, "The way Syrian refugees in Turkey use media: Understanding "connected refugees" through a non-media-centric and local approach," *Communications,* vol. 43, pp. 113--123, 2018.

[23] S. Silm and R. a. M. V. Ahas, "Are younger age groups less segregated? Measuring ethnic segregation in activity spaces using mobile phone data," *Journal of Ethnic and Migration Studies,* vol. 44, pp. 1797--1817, 2018.

[24] M. N. Ahmed, G. Barlacchi, S. Braghin, F. Calabrese, M. Ferretti, V. Lonij, R. Nair, R. Novack, J. Paraszczak and A. S. Toor, "A Multi-Scale Approach to Data-Driven Mass Migration Analysis," in *SoGood@ ECML-PKDD*, 2016.

[25] V. K. Singh, B. Bozkaya and A. Pentland, "Money walks: implicit mobility behavior and financial well-being," *PloS one,* vol. 10, p. e0136628, 2015.

[26] D. Gundogdu, O. D. Incel, A. A. Salah and B. Lepri, "Countrywide arrhythmia: emergency event detection using mobile phone data," *EPJ Data Science,* vol. 5, no. 1, p. 25, 2016.

[27] G. Song, L. a. Y. W. Yu and M. Xiujun, "Discovering Spatial Interaction Communities from Mobile Phone Data," *Transaction in GIS,* pp. 463-481, 2013.

[28] L. Fiorio, G. Abel, J. Cai, E. Zagheni, I. Weber and G. Vinue, "Using twitter data to estimate the relationship between short-term mobility and long-term migration," *Proceedings of the 2017 ACM on Web Science Conference,* pp. 103--110, 2017.

[29] F. Hubl, S. Cvetojevic, H. Hochmair and G. Paulus, "Analyzing refugee migration patterns using geo-tagged tweets," *ISPRS International Journal of Geo-Information,* p. 302, 2017.

[30] F. Wu, M. Zhu, X. Zhao, Q. Wang, W. Chen and R. Maciejewski, "Visualizing the time-varying crowd mobility," in *SIGGRAPH Asia 2015 Visualization in High Performance Computing*, 2015.

[31] H. Barbosa, M. Barthelemy, G. Ghoshal, C. R. James, M. Lenormand, T. Louail, R. Menezes, J. J. Ramasco, F. Simini and M. Tomasini, "Human mobility: Models and applications," *Physics Reports,* vol. 734, pp. 1-74, 2018.

[32] C. Song, Z. Qu, N. Blumm and A. L. Barabási, "Limits of predictability in human mobility," *Science,* vol. 327, pp. 1018-1021,

2010.

[33] L. Alessandretti, P. Sapiezynski, V. Sekara, S. Lehmann and A. Baronchelli, "Evidence for a conserved quantity in human mobility," *Nature Human Behaviour,* vol. 2, p. 485–491, 2018.

[34] M. C. Gonzalez, C. A. Hidalgo and A. L. Barabasi, "Understanding individual human mobility patterns," *Nature,* vol. 453, p. 779, 2008.

[35] T. Kevin and D. Matt, "Trajectory similarity measures," *SIGSPATIAL Special,* vol. 7, pp. 43-50, 2015.

[36] M. Lenormand, A. Bassolas and J. J. Ramasco, "Systematic comparison of trip distribution laws and models," *Journal of Transport Geography,* vol. 51, pp. 158-169, 2016.

[37] G. Yuan, P. Sun, J. Zhao, D. Li and C. Wang, "A review of moving object trajectory clustering algorithms," *Artificial Intelligence Review,* pp. 123--144, 2017.

[38] G. Atluri, A. Karpatne and V. Kumar, "Spatio-temporal data mining: A survey of problems and methods," *ACM Computing Surveys (CSUR),* p. 83, 2018.

[39] P. Barsocchi, M. G. Cimino, E. Ferro, A. Lazzeri, F. Palumbo and G. Vaglini, "Monitoring elderly behavior via indoor position-based stigmergy," *Pervasive and Mobile Computing,* pp. 26--42, 2015.

[40] A. L. Alfeo, M. G. C. A. Cimino, S. Egidi, B. Lepri and G. Vaglini, "A stigmergy-based analysis of city hotspots to discover trends and anomalies in urban transportation usage", IEEE Transactions on Intelligent Transportation Systems, IEEE, Vol. 19, Issue 7, Pages 2258-2267, 2018, (ISSN 1524-9050).

[41] V. Nee and R. Alba, "Rethinking assimilation theory for a new era of immigration," *The new immigration,* pp. 49--80, 2012.

[42] A. Lifanova, H. Y. Ngan, A. Okunewitsch, S. Rahman, S. Guzmán, N. Desai, M. Özsari, J. Rosemeyer, R. Pleshkanovska, A. Fehler and others, "New Locals: Overcoming Integration Barriers with Mobile Informal and Gamified Learning," in *Proceedings of the International Conference on Information Communication Technologies in Education*, 2016.

[43] F. A. Clark, "The concepts of habit and routine: A preliminary theoretical synthesis," *The Occupational Therapy Journal of Research,* vol. 20, no. 1_suppl, pp. 123S--137S, 2000.

[44] T. Jansen, N. Chioncel and H. Dekkers, "Social cohesion and integration: Learning active citizenship," *British Journal of Sociology of Education,* vol. 27, no. 02, pp. 189-205, 2006.

[45] A. Almaatouq, F. Prieto-Castrillo and A. Pentland, "Mobile communication signatures of unemployment," in *International conference on social informatics*, 2016.

[46] M. Bramer, F. Coenen and M. Petridis, "Research and Development in Intelligent Systems," in *AI-2007, the Twenty-seventh SGAI International Conference on Innovative Techniques and Applications of Artificial Intelligence*, 2007.

[47] C. Licoppe, D. Diminescu, Z. Smoreda and C. Ziemlicki, "Using mobile phone geolocalisation for 'socio-geographical'analysis of co-ordination, urban mobilities, and social integration patterns," *Tijdschrift voor economische en sociale geografie,* vol. 99, no. 5, pp. 584-601, 2008.

[48] A. Hebbani, V. Colic-Peisker and M. Mackinnon, "Know thy neighbour: Residential integration and social bridging among refugee settlers in Greater Brisbane," *Journal of Refugee Studies,* vol. 31, no. 1, pp. 82-103, 2017.

[49] M. Cimino, A. Lazzeri and G. Vaglini, "Improving the analysis of context-aware information via marker-based stigmergy and differential evolution," in *International Conference on Artificial Intelligence and Soft Computing (ICAISC 2015)*, Zakopane, Poland, 2015.

[50] L. Marsh and C. Onof, "Stigmergic epistemology, stigmergic cognition," *Cognitive Systems Research,* vol. 9, no. 1-2, pp. 136-149, 2008.

[51] D. Vernon, G. Metta and G. Sandini, "A survey of artificial cognitive systems: Implications for the autonomous development of mental capabilities in computational agents," *IEEE transactions on evolutionary computation,* vol. 11, no. 2, pp. 151-180, 2007.

[52] A. Alfeo, P. Barsocchi, M. Cimino, D. La Rosa, F. Palumbo and G. Vaglini, "Sleep behavior assessment via smartwatch and stigmergic receptive fields," *Personal and Ubiquitous Computing,* vol. 22, no. 2, pp. 227-243, 2017.

[53] A. Alfeo, M. Cimino, S. Egidi, B. Lepri, A. Pentland and G. Vaglini, "Stigmergy-based modeling to discover urban activity patterns from positioning data", in proceedings of "Social, Cultural, and Behavioral Modeling: 10th International Conference", (SBP-BRiMS 2017), vol. 10354 LNCS, p.p. 292-302. Springer. Washington, DC, USA, July 5-8, 2017.

[54] A. L. Alfeo, M. G. C. A. Cimino, B. Lepri and G. Vaglini, "Detecting Permanent and Intermittent Purchase Hotspots via Computational Stigmergy," in *The 8th International Conference on Pattern Recognition Applications and Methods (ICPRAM 2019)*, 2019. Doi: 10.5220/0007581308220829.

[55] K. S. Kung, K. Greco, S. Sobolevsky and C. Ratti, "Exploring universal patterns in human home-work commuting from mobile phone data," *PloS one,* vol. 9, no. 6, p. e96180, 2014.

[56] L. Alexander, S. Jiang, M. Murga and M. C. González, "Origin--destination trips by purpose and time of day inferred from mobile phone data," *Transportation research part c: emerging technologies,* vol. 58, pp. 240-250, 2015.

[57] X. V. Del Carpio and M. Wagner, The impact of Syrians refugees on the Turkish labor market, The World Bank, 2015.

[58] B. Balkan and S. Tumen, "Immigration and prices: quasi-experimental evidence from Syrian refugees in Turkey," *Journal of Population Economics,* vol. 29, no. 3, pp. 657-686, 2016.

[59] M. Fawaz, "Planning and the refugee crisis: Informality as a framework of analysis and reflection," *Planning Theory,* vol. 16, no. 1, pp. 99-115, 2017.

[60] G. Chen, S. Hoteit, A. C. Viana, M. Fiore and C. Sarraute, "Enriching sparse mobility information in Call Detail Records," *Computer Communications,* vol. 122, pp. 44-58, 2018.

[61] O. Burkhard, R. Ahas, E. Saluveer and R. Weibel, "Extracting regular mobility patterns from sparse CDR data without a priori assumptions," *Journal of Location Based Services,* vol. 11, no. 2, pp. 78-97, 2017.

[62] E. McIsaac, Nation building through cities: A new deal for immigrant settlement in Canada, Caledon Institute of Social Policy, 2003.

[63] A. Llorente, M. Garcia-Herranz, M. Cebrian and E. Moro, "Social media fingerprints of unemployment," *PloS one,* vol. 10, no. 5, p. e0128692, 2015.

[64] L. Bakker, S. Y. Cheung and J. Phillimore, "The asylum-integration paradox: Comparing asylum support systems and refugee integration in the Netherlands and the UK," *International Migration,* vol. 54, no. 4, pp. 118-132, 2016.

[65] M. De Nadai, R. L. Vieriu, G. Zen, S. Dragicevic, N. Naik, M. Caraviello, C. A. Hidalgo, N. Sebe and B. Lepri, "Are safer looking neighborhoods more lively?: A multimodal investigation into urban life.," in *In Proceedings of the 24th ACM international conference on Multimedia*, 2016.

[66] A. Madanipour, G. Cars and J. Allen, Social exclusion in European cities: processes, experiences, and responses, vol. 23, Psychology Press, 2000.

[67] E. Lyytinen, "Refugees' Conceptualizations of "Protection Space": Geographical Scales of Urban Protection and Host--Refugee Relations," *Refugee Survey Quarterly,* vol. 34, no. 2, pp. 45-77, 2015.

[68] *One killed in brawl between locals and Afghan, Syrian migrants in Istanbul,* 2018.

[69] *Istanbul police evacuate refugees after clashes with locals,* 2018.

[70] M. D. Ward, A. Beger, J. Cutler, M. Dickenson, C. Dorff and B. Radford, "Comparing GDELT and ICEWS event data," *Analysis,* vol. 21, no. 1, pp. 267-297, 2013.

[71] C. Montag, K. Blaszkiewicz, R. Sariyska, B. Lachmann, I. Andone, B. Trendafilov, M. Eibes and A. Markowetz, "Smartphone usage in the 21st century: who is active on WhatsApp?," *BMC research notes,* vol. 8, no. 1, p. 331, 2015.

[72] G. Ranjan, H. Zang, Z.-L. Zhang and J. Bolot, "Are call detail records biased for sampling human mobility?," *ACM SIGMOBILE Mobile Computing and Communications Review,* vol. 16, no. 3, pp. 33-44, 2012.

[73] J. Wiese, J.-K. Min, J. I. Hong and J. Zimmerman, "You never call, you never write: Call and sms logs do not always indicate tie strength," in *18th ACM conference on computer supported cooperative work & social computing*, 2015.

[74] L. Pappalardo, M. Vanhoof, L. Gabrielli, Z. Smoreda, D. Pedreschi and F. Giannotti, "An analytical framework to nowcast well-being using mobile phone data," *International Journal of Data Science and Analytics*, pp. 75-92, 2016.

[75] A. M. Tompkins and N. McCreesh, "Migration statistics relevant for malaria transmission in Senegal derived from mobile phone data and used in an agent-based migration model.," *Geospatial health*, vol. 11, no. 1 Suppl, p. 408, 2016.

[76] V. K. Singh, B. Bozkaya and A. Pentland, "Money walks: implicit mobility behavior and financial well-being," *PloS one*, vol. 10, no. 8, p. e0136628, 2015.

[77] S. Silm and R. Ahas, "Ethnic differences in activity spaces: A study of out-of-home nonemployment activities with mobile phone data," *Annals of the Association of American Geographers*, vol. 104, no. 3, pp. 542-559, 2014.

[78] S. Niwattanakul, J. Singthongchai, E. Naenudorn and S. Wanapu, "Using of Jaccard coefficient for keywords similarity," in *Proceedings of the International MultiConference of Engineers and Computer Scientists*, 2013.

[79] A. Madanipour, "Social exclusion and space," in *The city reader*, Routledge, 2015, pp. 237-245.

[80] J. Karakoç and F. Do\ugruel, "The Impact of Turkey's Policy toward Syria on Human Security," *Arab Studies Quarterly*, vol. 37, no. 4, pp. 351-366, 2015.

[81] Y. Dong, F. Pinelli, Y. Gkoufas, Z. Nabi, F. Calabrese and N. V. Chawla, "Inferring unusual crowd events from mobile phone call detail records," in *Joint European conference on machine learning and knowledge discovery in databases*, 2015.

[82] M. Boustani, E. Carpi, H. Gebara and Y. Mourad, "Responding to the Syrian crisis in Lebanon Collaboration between aid agencies and local governance structures," *Refugee Research and Policy in the Arab World*, 2016.

[83] O. Altindag and N. Kaushal, "Do refugees impact voting behavior in the host country? Evidence from Syrian Refugee inflows in Turkey," 2017.

[84] *Kurdish militants claim responsibility for Turkey tunnel attack*, 2018.

[85] *130 migrants intercepted in Aegean heading for Chios*, 2018.

[86] "130 migrants intercepted in Aegean heading for Chios.," 7 3 2017. [Online]. Available: http://www.ansamed.info/ansamed/en/news/nations/greece/2017/03/06/130-migrants-intercepted-in-aegean-heading-for-chios_b60eec09-29a6-42a6-b846-3e747f71f000.html. [Accessed 13 7 2018].

[87] "Kurdish militants claim responsibility for Turkey tunnel attack.," 14 4 2017. [Online]. Available: https://www.reuters.com/article/us-turkey-blast/kurdish-militants-claim-responsibility-for-turkey-tunnel-attack-idUSKBN17E0MW. [Accessed 13 7 2018].

[88] "One killed in brawl between locals and Afghan, Syrian migrants in Istanbul.," 16 5 2017. [Online]. Available: http://www.hurriyetdailynews.com/one-killed-in-brawl-between-locals-and-afghan-syrian-migrants-in-istanbul-113111. [Accessed 13 7 2018].

[89] "Istanbul police evacuate refugees after clashes with locals.," 17 5 2017. [Online]. Available: https://www.trtworld.com/turkey/istanbul-police-evacuate-300-refugees-after-clashes-with-locals-358027. [Accessed 13 7 2018].

[90] Y. Ma, T. Lin, Z. Cao, C. Li, F. Wang and W. Chen, "Mobility viewer: An Eulerian approach for studying urban crowd flow," *IEEE Transactions on Intelligent Transportation Systems*, vol. 17, no. 9, pp. 2627-2636, 2016.

[91] C. Bauckhage, R. Sifa, A. Drachen, C. Thurau and F. Hadiji, "Beyond heatmaps: Spatio-temporal clustering using behavior-based partitioning of game levels," in *Computational Intelligence and Games (CIG), 2014 IEEE conference on*, 2014.

[92] B. Hugo, B. Marc, G. Gourab, C. R.James, L. Maxime, L. Thomas, M. Ronaldo, J. R. José, S. Filippo and T. Marcello, "Human mobility: Models and applications," *Physics Reports*, vol. 734, pp. 1-74, 2018.

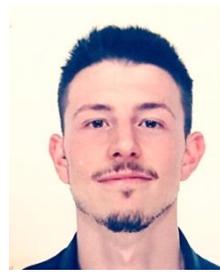

**Antonio Luca Alfeo** is a Postdoc research fellow at the Department of Information Engineering of the University of Pisa. He received the Ph.D. degree from the Ph.D. Program in Smart Computing with a thesis addressing the design of bioinspired approaches for machine learning and data analysis. In 2018 he was a visiting Ph.D. student at the Media Lab of the Massachusetts Institute of Technology. His research interests include the design of machine learning pipelines with applications in the context e-health, smart mobility, and industry 4.0.

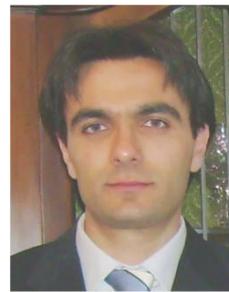

**Mario G.C.A. Cimino** is with the Department of Information Engineering (University of Pisa) as an Associate Professor. He is also a research associate at the Institute for Informatics and Telematics (IIT) of the Italian National Research Agency (CNR). In 2006, he was a visiting Ph.D. student at the Electrical and Computer Engineering Research Facility of the University of Alberta, Canada. In 2007, he received the Ph.D. degree in Information Engineering from the University of Pisa. His research focus lies in the areas of Swarm Intelligence and Business/Social Process Analysis, with particular emphasis on Stigmergic Computing, Workflow Mining and Simulation. He is (co-) author of more than 50 publications.

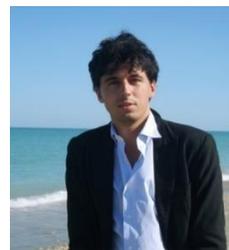

**Bruno Lepri** leads the Mobile and Social Computing Lab (MobS) at Bruno Kessler Foundation (Trento, Italy). Bruno is also Head of Research of Data-Pop Alliance. In 2010 he won a Marie Curie Cofund post-doc fellowship and he has held post-doc positions at FBK and at MIT Media Lab. He holds a Ph.D. in Computer Science from the University of Trento. His research interests include computational social science, personality computing, urban computing, network science, machine learning, and new models for personal data management and monetization. His research has received attention from several international press outlets and obtained the James Chen Annual Award for best UMUAI paper and the best paper award at ACM Ubicomp 2014. His work on personal data management was one of the case studies discussed at the World Economic Forum.

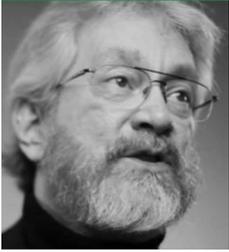
**Alexander "Sandy" Pentland** directs the MIT Connection Science and Human Dynamics labs and previously helped create and direct the MIT Media Lab and the Media Lab Asia in India. Forbes recently declared him one of the "7 most powerful data scientists in the world" along with Google founders and the Chief Technical Officer of the United States. He is a founding member of advisory boards for Google, AT&T, Nissan, and the UN Secretary General, a serial entrepreneur who has co-founded more than a dozen companies including social enterprises such as the Data Transparency Lab, the Harvard-ODI-MIT DataPop Alliance and the Institute for Data Driven Design. He is a member of the U.S. National Academy of Engineering and leader within the World Economic Forum.

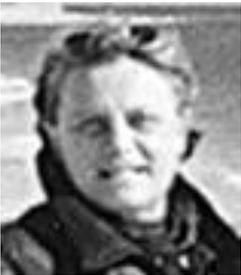
**Gigliola Vaglini** was born in Pisa, Italy, on June 11, 1952. She received the M.S. degree in Computer Science from the University of Pisa. She was a research assistant at the University of Pisa, Department of Computer Science, an associate Professor at the University of Naples, Federico II, Department of Mathematics, and then at the University of Pisa, Department of Information Engineering; from 2002 she is a Full Professor at the same Department. Her research has been concerned mainly with formal methods for specification and verification of concurrent and distributed systems, in particular the model checking of concurrent systems. More recently her research activity was focused on the use of the stigmergic paradigm to aggregate data supplied by different sources and to obtain a distributed control on autonomous systems.